\documentclass[conference]{IEEEtran}
\IEEEoverridecommandlockouts
\usepackage{cite}
\usepackage{amsmath,amssymb,amsfonts}
\usepackage{algorithmic}
\usepackage{graphicx}
\usepackage{textcomp}
\usepackage{url}
\usepackage{xcolor}
\def\BibTeX{{\rm B\kern-.05em{\sc i\kern-.025em b}\kern-.08em
    T\kern-.1667em\lower.7ex\hbox{E}\kern-.125emX}}
\begin{document}

\title{Games! What are they good for? \\ The Struggle of Serious Game Adoption for Rehabilitation}

\author{\IEEEauthorblockN{
Maria Micaela Fonseca\IEEEauthorrefmark{1}, 
Nuno Fachada\IEEEauthorrefmark{2}\IEEEauthorrefmark{3}, 
Micael Sousa\IEEEauthorrefmark{4}, 
Jorge Oliveira\IEEEauthorrefmark{1},\\
Pedro Rodrigues\IEEEauthorrefmark{1},
Sara Sousa\IEEEauthorrefmark{5},
Cláudia Quaresma\IEEEauthorrefmark{6},
and
Phil Lopes\IEEEauthorrefmark{1} 
}

\IEEEauthorblockA{\IEEEauthorrefmark{1}Lusófona University, HEI‐Lab: Digital Human‐Environment Interaction Labs, \\
Campo Grande, 376, 1749-024 Lisboa, Portugal
}

\IEEEauthorblockA{\IEEEauthorrefmark{2}Lusófona University, ECATI, 
Campo Grande, 376, 1749-024 Lisboa, Portugal}

\IEEEauthorblockA{\IEEEauthorrefmark{3}Center of Technology and Systems (UNINOVA-CTS) and Associated Lab of Intelligent Systems (LASI),\\  2829-516 Caparica, Portugal }

\IEEEauthorblockA{\IEEEauthorrefmark{4}
School of Architecture, Planning and Environmental Policy, University College Dublin, \\ D04 V1W8 Dublin, Ireland}

\IEEEauthorblockA{\IEEEauthorrefmark{5}Centro de Estudos Comparatistas, Faculdade de Letras, University of Lisbon, \\
Alameda da Universidade, 1600-214 Lisboa, Portugal}

\IEEEauthorblockA{\IEEEauthorrefmark{6}LIBPhys-UNL, Physics Department, NOVA School of Science and Technology, NOVA University Lisbon \\
2829-516 Caparica, Portugal}

\thanks{Corresponding author: Maria Micaela Fonseca (e-mail: micaela.fonseca@ulusofona.pt)}
}

\maketitle

\begin{abstract}
The field of serious games for health has grown significantly, demonstrating effectiveness in various clinical contexts such as stroke, spinal cord injury, and degenerative neurological diseases. Despite their potential benefits, therapists face barriers to adopting serious games in rehabilitation, including limited training and game literacy, concerns about cost and equipment availability, and a lack of evidence-based research on game effectiveness.  Serious games for rehabilitation often involve repetitive exercises, which can be tedious and reduce motivation for continued rehabilitation, treating clients as passive recipients of clinical outcomes rather than players. This study identifies gaps and provides essential insights for advancing serious games in rehabilitation, aiming to enhance their engagement for clients and effectiveness as a therapeutic tool. Addressing these challenges requires a paradigm shift towards developing and co-creating serious games for rehabilitation with therapists, researchers, and stakeholders. Furthermore, future research is crucial to advance the development of serious games, ensuring they adhere to evidence-based principles and engage both clients and therapists. This endeavor will identify gaps in the field, inspire new directions, and support the creation of practical guidelines for serious games research.

\end{abstract}


\section{Introduction}

The widespread presence and familiarity with technology (analogue and digital) and the advancements of game design have made the fulfillment of serious games more propitious. Serious games appear to be one of the most appealing interactive media for empowering clients and providing tailored experiences for health treatments, with significant growth \cite{laamarti2014overview}.
Serious games emerged as a promising tool for a wide range of healthcare applications \cite{dorner2016serious}, including healthcare education and training, monitoring, diagnosis, prevention, and rehabilitation. Using these games to enhance healthcare outcomes has generated considerable interest among a growing community of researchers, game designers, game developers, and healthcare professionals seeking innovative solutions to address complex health challenges \cite{abd2022artificial}.

Based on previous works’ insights, game-based therapy significantly improved and yielded better functional outcomes for motor and cognitive rehabilitation \cite{amorim2020serious,laver2017virtual}.
While the findings are promising, they primarily concentrate on clinical viewpoints. Most available prototypes (or games) do not possess the characteristics required for sustained engagement as a long-term serious game, often employed by clients independently and enjoyably.

Furthermore, the research community primarily emphasises introducing novel technologies and different applications to enhance serious games for healthcare while neglecting critical aspects such as structured evidence and clear consensus regarding the development of engaging games, despite the type of game and platform being used. Additionally, there is often a disregard for the significant and noteworthy connection between the strength of games and the collaborative involvement of clients and healthcare professionals in the co-creation process \cite{marsilio2021co}.

This study seeks to identify existing gaps and elucidate the process by offering essential elements for shifting the paradigm of serious games employed in rehabilitation. Ultimately, this study aims to contribute to the challenge of identifying how to make serious games for health more engaging for clients and a proper therapeutic tool for health professionals. Despite the large breadth of existent work, there is still a struggle to make the leap from laboratory testing to a fully usable system in the field. Knowing how to develop, choose and adapt game based approaches for rehabilitation would be a game-changing quest.

This paper is organized as follows. Section~\ref{sec:transfrehab}, ``Serious games: Transforming Rehabilitation'', characterizes various approaches to serious games and provides examples of VR and board games in rehabilitation, emphasizing their widespread use and benefits. It also observes that most existing games are predominantly gamified systems centered around simulated activities or single tasks.
Section~\ref{sec:unlocking}, ``Unlocking the True Potential: Exploring Therapists' Roles'', addresses remote therapy and the barriers to redefine serious games. Section~\ref{sec:multidis}, ``Multidisciplinarity and the role of game design in rehabilitation games'', elaborates on the significance of playful experiences within serious games, emphasizing that the quality of game design and feedback directly influence stakeholders' adoption. Section~\ref{sec:fiction}, ``Importance of Fiction in Rehabilitation Games'', details the importance of storytelling, leveraging the power of narrative used in serious games for health. Finally, Section~\ref{sec:conclusion}, ``Conclusion: the "abandoned" artefact'', explores insights for the future by providing knowledge about how to develop more effective serious games and general takeaways for other researchers.

\section{Serious games: Transforming rehabilitation}
\label{sec:transfrehab}

The term \emph{serious games} is often used as a "catch-all" for games that serve a purpose beyond just entertainment, where the goal is to leverage the advantages of play for applications in education, self-motivation, health, and beyond. Rehabilitation can be defined as the process of restoring and improving the physical, mental, and social functioning of individuals who have been affected by injury or disease. This process involves a coordinated effort by healthcare professionals, patients, and their families to maximize the individual's abilities and help them achieve their full potential in their environment \cite{who}.
Given the rise of alternative input devices that leverage physical movement for game input control (e.g., Wii Controller, Wii Fit, Xbox Kinect, Index Controller), it is unsurprising that current literature has slowly shifted towards such devices and the games that use them, for its potential for rehabilitative-focused serious games.  It is commonly assumed that games, and more recently video games, have the capacity to engage and motivate audiences, thereby potentially enhancing the memorability and effectiveness of the experience when applied in serious contexts \cite{silva2023soft,wang2016systematic}. Motivation plays a crucial role in rehabilitation as it is often associated with better therapeutic outcomes \cite{fleming2017serious}. Nevertheless, describing motivation can be complex, as it encompasses a multifaceted concept \cite{vahlo2019five}. It can be perceived as a psychological attribute that propels a client towards the initiation and/or sustenance of goal-directed behaviour.
A motivated client is willing to exert effort without requiring excessive encouragement or expressing dissatisfaction with treatment demands. Intrinsic motivation arises when clients find interest, enjoyment, and satisfaction in an activity, and playing games inherently taps into this intrinsic motivation \cite{drummond2017serious}. Motivation fosters engagement, ensuring the ongoing utilization of the game as a therapeutic instrument that delivers enjoyment to the user. Playing the game continuously or from start to finish is a common challenge to game designer. 
Several serious games are applied to healthcare in academia and industry contexts with various formats depending on their therapeutic modalities \cite{abd2022artificial}, such as exergames \cite{xu2020results}, commercial games \cite{sawa2022potential}, \cite{choi2014effectiveness}, digital-based interventions \cite{banos2022current}, the integration of biosensors \cite{kothgassner2022virtual} and haptic feedback, (analogue) board/tabletop games \cite{gauthier2019board}, and virtual reality (VR) \cite{corbetta2015rehabilitation}. The rapid pace of innovation provides additional challenges for the serious games community. Rather than providing an overview of the extensive range of games available for rehabilitation, this study will concentrate on two modalities with varying levels of popularity. Firstly, the study will examine board games (Subsection~\ref{sec:transfrehab:bg}), which have received limited attention in research literature yet possess a key advantage in promoting social interactions. Secondly, the investigation will focus on VR (Subsection~\ref{sec:transfrehab:vr}), a technology that has gained significant popularity in healthcare. Despite being two completely different types of games, analogue games can work as a strategy to gradually introduce game based approaches for users less exposed to digital games. Also, analogue games are a proven way to prototype and establish co-design processes for digital game development (including VR games) \cite{ham2015tabletop}. Embracing and exploring this combination can help discover novel strategies for engaging clients in serious games.
\subsection{Motivation}
\label{sec:transfrehab:motiv}

Self-Determination Theory (SDT) provides a foundational framework for understanding motivational mechanisms, particularly in behavior change. This theory offers a conceptual basis to examine how contextual, social-psychological, and individual factors influence human behavior. It is central to the study of sustained behavior change in rehabilitation settings \cite{ryan2000intrinsic, terroni2015association}. SDT, developed by Ryan and Deci \cite{ryan2000intrinsic}, identifies three basic psychological needs---competence, autonomy, and relatedness---as essential drivers of intrinsic motivation.

Competence refers to the need to feel effective in performing challenging tasks. For example, when individuals successfully achieve rehabilitation milestones, their self-efficacy is reinforced, which in turn motivates them to persist with the intervention. Autonomy is the need to feel a sense of control over one's behaviors and decisions. Rehabilitation programs that adapt exercises to individual needs and preferences foster autonomy, empowering individuals to take ownership of their recovery journey. Lastly, relatedness involves a sense of connection or belonging. When individuals feel cared for and supported by healthcare professionals, they are more likely to engage actively and remain committed to their recovery \cite{ryan2000intrinsic}.

In the broader context of behavior change technologies (BCTs), Optimal Functioning Theory (OIT) complements SDT by focusing on sustaining newly acquired behaviors. However, many existing BCTs overlook the importance of fostering sustainable self-regulation, which is critical for long-term success. The transition from identified regulation (behavior adopted due to perceived importance) to integrated regulation (behavior fully aligned with one’s core values) is a promising approach for achieving durable behavior change, particularly in managing chronic conditions \cite{gerstenberg2023designing}. Integrated regulation requires less effort, enhances satisfaction, and promotes well-being, making it vital for self-management.

Equi, a theory-driven functional prototype, integrates SDT principles to facilitate this transition. By emphasizing competence, autonomy, and relatedness, Equi aims to enhance intrinsic motivation and support sustainable behavior change. Grounded in SDT, the prototype fosters autonomous motivation and perceived competence, which is critical for meaningful and lasting change \cite{tyack2024self}. Moreover, it addresses the depletion of self-regulation resources, a common barrier in behavior change efforts, by fostering social contexts that support autonomy rather than imposing control or demotivation \cite{srisodsaluk2023application}.

Integrating SDT principles into serious games provides a user-centered approach that enhances engagement through interactive and meaningful gameplay. By embedding strategies that prioritize competence, autonomy, and relatedness, serious games can foster intrinsic motivation, encouraging sustained participation and deeper connection with desired behavioral outcomes. This approach not only improves user adherence and satisfaction but also supports long-term behavior change and well-being, leveraging the immersive and adaptive potential of serious games.

\subsection{Board Games for Health}
\label{sec:transfrehab:bg}

Board games have a longstanding history of being utilised in therapeutic contexts. At least since the 1950s, there are literature references about researchers using games for therapeutic uses, mostly in psychotherapy and for children \cite{stone2019game}. However, the playful formal therapeutic approaches can be traced back to the beginning of the 20th century \cite{o2015handbook}. It is crucial to distinguish between play and games as they are distinct concepts. Moreover, delving deeper into history, the emergence of Parlour Games designed for adults in the XVIII and XIX centuries exemplifies the popularity of board games that served specific purposes and aligned with particular social contexts. Some of these games were created by companies aiming for families, trying to transmit morals and ethics in Western societies (e.g., Victorian society). Playing "The Game of Life" by Milton Bradley was one example of these roll-and-move games \cite{woods2012eurogames}. These were not therapeutic games in the contemporary sense, but they transmitted values. The same for the military, which has used modern simulation games since the XIX century as training tools (e.g., Kriegsspiel) \cite{smith2010long}. 

Board games as learning spaces are prevalent in various fields, simulating real-world scenarios and facilitating the understanding of the subject matter. Research on board games primarily focuses on those designed for educational purposes \cite{bayeck2020examining}. Similar to other game-based approaches, using analogue games for health aims to enhance healthcare education by fostering the acquisition of knowledge and skills \cite{gauthier2019board}. Bochennek et al.\cite{bochennek2007more} conducted a review on card and board games used for medical teaching, addressing game design mechanisms, components (such as win conditions and feedback cycles), target audience, and suitable gameplay settings. The authors also explore knowledge transfer and the evaluation of outcomes. However, the authors acknowledge that their proposal does not guarantee that the games are fun. But if fun is not a requirement, what is the difference between using games or any other therapeutic technique? This lack of understanding is a key factor to understand what motivate users to actively participate in therapy. As a result, there is a growing trend towards adapting entertainment games for therapeutic purposes \cite{bayeck2020examining,rosamonitoring}. Taking into account the implementation of rehabilitation games in real-world clinical settings, Jung et al.~\cite{jung2020rehabilitation} stated that clients report that board games commonly used in clinical settings were childish (due to the use of cartoon-like imagery), leading to clients’ negative perceptions towards serious games. The balance between entertainment and seriousness is not yet solved in games applied to health. 

One significant advantage of using board games is their capacity to foster face-to-face interactions. This feature would bring together clients, family members, caregivers and even therapists trough the same media. This aspect has not received adequate attention or consideration in the existing literature. \cite{calleja2022unboxed, booth2021board}

Recently, board game publishers have been trying to market their games as therapeutic and/or learning tools, aiming for new social impacts. For example, Devir created the Devir Home Academy \cite{devirhac}, offering tips and support material to use their games as serious games. Mercurio games funded PhD industrial grants to analyse their games and find what is suited to specific skills \cite{vita2022cognitive}. There is also the recent case of Access+ line from Asmodee \cite{asmodeeal}, where commercial games that deal with cognitive skills were adapted to fit users needs like disability or other limitations. This practical knowledge transfer can leverage the available tools for therapists, although more structured methods are required. Through a systematic classification of games and gamme-based approaches, practitioners gain the ability to select and participate in the development of games that align with specific therapeutic contexts.

\subsection{Virtual reality-based interventions}
\label{sec:transfrehab:vr}

In recent years, VR-based interventions in clinical settings have witnessed a significant upsurge. These innovative approaches allow users to engage with immersive environments that closely resemble real-life experiences \cite{weiss2006virtual}. 
Within this realm, research suggests that VR environments have proven effective in relaxing \cite{gamito2023relaxing} and reducing stress and anxiety \cite{diemer2014virtual}. Moreover, VR has demonstrated successful applications in healthcare contexts, including pain management \cite{matsangidou2017clinical}, social anxiety \cite{premkumar2021effectiveness}, mental health disorders  \cite{kothgassner2023virtual}, such as post-traumatic stress disorder \cite{vianez2022virtual},  \cite{kim2022effectiveness}, autism spectrum disorder \cite{bravou2022applications}, depression \cite{migoya2020feasibility},  cognitive and motor rehabilitation for stroke  \cite{chen2022effects}, \cite{chen2022effectiveness}, \cite{lee2019effects}, \cite{kowsalya2024rewiring} emotion regulation  \cite{colombo2021virtual}, and phobias \cite{demir2023efficacy}. Previous research has extensively examined the comprehensive benefits associated with VR/game-based therapy \cite{kim2005swot}.

VR promotes the rehabilitation of cognitive function and recovery of daily living activities (ADL) in patients with post-stroke cognitive impairment. Consequently, VR can be a valuable complementary approach to conventional cognitive interventions \cite{chen2022effects}.
Taking exposure therapy as an example \cite{steinman2016exposure}, VR-based exposure therapy (VRET) employs computer simulations to elicit comparable reactions to real-life phobic encounters, thereby emphasizing its robust ecological validity  \cite{morina2015can}. 
Despite its benefits, mental health providers have not yet embraced VRET for widespread use \cite{rizzo2017clinical}.
Recently, VR emerges as a promising tool in neurorehabilitation, enhancing conventional therapies and potentially reducing hospitalization duration \cite{maggio2024virtual}.
 
Although VR gained tremendous interest among researchers and healthcare professionals, previous research studies have offered limited insights into the development process of VR/game-based therapy, provided inadequate information regarding real-world and clinical settings, evaluations of their effectiveness, and frequently referenced poorly designed research studies \cite{kato2013role}.

Nonetheless, despite the ecological validation, the current state of VR applications often demonstrates simplistic characteristics, primarily focusing on developing first-person virtual experiences \cite{tao2021immersive}, replicating real-life activities or single tasks. As a result, there is a missed opportunity to fully leverage the potential of game design principles and narrative elements and empower clients through agency and autonomy.

\section{Unlocking the True Potential: Exploring Therapists' Roles }
\label{sec:unlocking}

The COVID-19 global pandemic and the subsequent social distancing guidelines triggered considerable unforeseen changes to the delivery of healthcare services. COVID-19 suddenly forced an across-the-board transition to remote therapy and telerehabilitation, likely to persist beyond the pandemic. 

The forced uptake of these remote services emerged as an opportunity to eliminate barriers and health professionals’ resistance to remote access, contributing to a general acknowledgement that innovative healthcare services will be an integral part of everyday practice. Healthcare professionals prepared themselves and their clients for the remote transition, and the pandemic led to a more positive mindset toward remote therapy \cite{bekes2020stretching}. Remote therapy opened the pathway for integrating home-based innovative solutions, thereby promoting clients’ engagement in essential rehabilitation routines. Furthermore, this new mindset allowed new solutions to serve as a complementary approach to conventional rehabilitation, providing an outstanding opportunity for the serious games community to take advantage. 

New guidelines and management are needed to enable the reconceptualization of remote therapy systems \cite{anil2021scope,bezuidenhout2022telerehabilitation,caldeira2021towards,ross2022staff}. The monitoring of remote digital solutions must be carefully designed following two considerations: 1) clients wish transparency and understanding of their progress; and, 2) health professionals need the information to support clinical decision-making, including clients’ adherence and progress. The importance of health professionals is supported by the concept of the therapeutic alliance, which has been considered an important predictor of treatment outcomes \cite{bordin1979generalizability,hall2010influence}. These outcomes can be explained by the client’s trust in the treatment agent, supported by a patient-therapist bond \cite{krasovsky2020will,moore2020therapeutic}. For remote therapy, digital therapeutic alliance (DTA) can be applied. Thus, it is reasonable to assume that difficulties on the part of health professionals in implementing a remote therapy program with new solutions may affect the therapeutic alliance and treatment outcomes. Effective communication strategies and easy-to-use interfaces were identified as key facilitators of successful services \cite{anil2021scope,ross2022staff}. However, what kind of needs clients and health professionals seek to meet their expectations are not well defined \cite{akinsiku2021s,caldeira2021towards}, and there are still some unmet challenges \cite{saaei2021rethinking}.  Improvements in data visualization can increase situational awareness and maximize the utility of client data \cite{o2018visualization}. Health professionals struggle to manage, analyze, interpret, and present the vast amount of data meaningfully. Human-computer interaction (HCI) can shape methods and knowledge that can be used to foster the DTA in digital remote solutions \cite{balcombe2022human,d2020digital,tong2022digital}.

Given the aforementioned issues, serious games have the potential to capitalize on this opportunity and assume a pivotal position.
Several studies have investigated the use of serious games in rehabilitation to enhance client engagement and motivation. For example, clinical studies with commercial games suggest that serious games can improve motor function recovery rates \cite{marques2021effectiveness}. It is recognized that clients' engagement and adherence levels vary in conventional rehabilitation, and therapists are crucial for maintaining engagement and ensuring exercise quality, providing additional feedback on clients' performance. Moreover, the current literature on serious games for rehabilitation suggests the possibility of heterogeneous engagement patterns among stroke patients, influenced by their cognitive and impairment conditions \cite{alankus2010towards}. Therefore, it is imperative to consider the client's individualized requirements. In contemporary times, the necessity to relinquish the one-size-fits-all approach is undeniable. Therapists acknowledge the significance of personalized approaches in client treatment and prepare strategic rehabilitation sessions to provide personalized, patient-centred therapy, and actively adjust their activities during the therapy sessions based on clients’ responses. Therefore, therapists fulfil comprehensive and orchestrating roles during therapy sessions to maximize clinical outcomes for their clients \cite{jung2020rehabilitation, ross2022staff,krasovsky2020will}. From game design perspective, the co-creation of the game with therapists fits the user-centered design approaches recommended by the pivoting literature \cite{{fullerton2014game,tekinbas2003rules,schell2008art}}. As stated before, analogue games can be the product (game to be applied) or the means to prototype digital games. 

The clinical outcomes and achievements outshine the engagement and the playful experience.
Therapists firmly believe that their continuous involvement is necessary to maintain control over each aspect of therapy sessions. This perspective often yields negative attitudes towards serious games and prevents the implementation of a balanced and adaptive game difficulty approach. Moreover, the concept of remote rehabilitation replacing clinicians evokes understandable apprehension among healthcare professionals. Convincing the healthcare community that modifying the rehabilitation approach does not compromise the quality of care and their jobs \cite{krasovsky2020will} is of the utmost importance. The willingness of therapists to adopt new innovative solutions is essential in ensuring the effectiveness of serious games implementation.

Extensive literature is available on the use of individualization in numerous serious games. Serious games face a unique challenge, as they are developed to address specific problems and not necessarily for a specific segment of the gaming community \cite{sajjadi2022individualization,carlier2023software}. To address this challenge, there have been efforts to implement adaptive difficulty in the serious gaming context. Hendrix et al.\cite{hendrix2018implementing} present a six-step plan for implementing dynamic difficulty adjustment based on the player's performance. Sajjadi et al.\cite{sajjadi2022individualization} offer a comprehensive overview of the player aspects employed for individualization in the field of serious games. 
Therapists can leverage the assessment of performance in serious games to monitor measurable parameters over time (taking advantage of game metrics), facilitating the optimal customization of difficulty levels for clients and goal establishment within the system.

Healthcare professionals must prioritize the development of meaningful connections with clients, irrespective of their mode of engagement, be it remote rehabilitation or in-person interactions. Serious games emerge as a noteworthy form of communication, facilitating robust therapist-client relationships and fostering favourable transformations in social behavior \cite{mayer2016utility,drummond2017serious}. To leverage the benefits of serious games, therapists should be literate in games and technology to enhance care accessibility, reduce expenses, and overcome geographical limitations, thus promoting equality. Therapists should collaborate with game development teams, and the games themselves should be sufficiently flexible to be adapted to specific therapy protocols during rehabilitation. These are considerable challenges, adding to the need to deliver engaging games.
Considering the development of these serious games it is recommended that therapists collaborate with the game development teams and that the games are flexible to be adapted (to clients and therapists' specific protocols) during rehabilitation. These are considerable challenges, adding to the need to deliver engaging games. 

\section{The role of game design and narrative in rehabilitation games}
\label{sec:multidis}

Literature surrounding rehabilitation games can significantly differ on which priorities are communicated and explored. For example, communities more focused on health tend to prioritize the efficacy of games as treatment and the technology itself~\cite{skjaeret2016exercise}, while game research-oriented communities tend to explore these games from a game design perspective, such as ``re-framing'' rehabilitation exercises into playful activities~\cite{pirovano2016exergaming}. 

Despite the breadth of research within the field, such communities still tend to have little to no interaction between each other, which in theory can leave some of the work explored ``incomplete'' or lacking certain rigour given the deficit of knowledge each expertise offers. This lack of communication is often propagated by specific research priorities that often do not compensate for multidisciplinary research or any type of long-term collaboration. Furthermore, the hassle of dealing with diverging languages that are typical of such communities can lend itself to frustrating collaboration~\cite{oborn2010knowledge}. 

Designing meaningful rehabilitation playing experiences can feel lacking due to the fact that the medium itself requires a wide range of skills that can optimize the playing experience, extending its lifespan, while maintaining the rigour and efficacy of an actual rehabilitative treatment. Thus, it is common that tools developed in such closed-off communities can often lead to a series of pitfalls that can reduce their probability of being used "en-masse" in the field.

A certain balance is necessary for the benefits of using games to "shine", as the core reasoning for using them is to motivate patients~\cite{skjaeret2016exercise,pirovano2016exergaming}. It is also important for designers to consider the state of mind of the individuals themselves, as certain exercises can be painful or require extensive concentration and energy. Thus, designing rehabilitation games requires a different approach and methodology than traditional game design, given that the output is not solely entertainment but should also be motivating and slowly push the patient towards their rehabilitative goals. Furthermore, in the perfect scenario such games should be able to capture the attention of the individual on the long-term, rather than the short-term given that rehabilitation can be slow and arduous process and motivating patients in these first few months (at the very least) can be critical to gain sufficient momentum to go through the entire process. 

Each type of game has its features and advantages when inspiring the development of serious games. Next, we will detail the game design features to be considered for analogue and digital games (VR) and how each informs the other. This approach will clarify the importance of game design to deliver therapeutic games and the need to establish a co-design approach from the design and therapeutics perspectives to deliver better serious games. Although they might seen as artificially separated, some commercial entertainment games are true hybrids, combining the physical components and interactions of a board game with the digital technologies of VR and a defined narrative for the players to explore. One of the cases is the popular game "Chronicles of Crime" from Lucky Duck Games. 

\subsection{Analogue games}
\label{sec:multidis:analog}

Analogue games, especially those that require dexterity, are used formally and informally as rehabilitation tools, sometimes even unconsciously, by therapists. Orthogames can be adapted by therapists and transformed into serious games that allow therapists to explore memory and motor skills using the game's pieces and mechanics and the users’ ability to develop storytelling \cite{rosamonitoring}. Inspired by the existing games, there are cases of developing simple analogue games as rehabilitation exercises \cite{rosa2022desempenho}. In these last cases, there is always the problem that the game might not be as engaging as one created for entertainment. This limitation is why some researchers are investing in modding existing analogue games since the game system is tested and proven to engage users \cite{estrada2021cognitive,rosamonitoring,vita2022cognitive}. There is the need to consider the player profiles \cite{zagalo2020engagement}, the types of experiences each game provides and the context/environment where the game is played \cite{martinho2023cssii}. All these dimensions can affect player experience and engagement and, in the case of therapeutic games, whether therapeutic objectives are reached. Overall, it is the type of experience of the analogue games that is at stake and is challenging, connecting the gameplay experiences to the therapeutic goals.

\subsection{Type of experience for the player?}
\label{sec:multidis:experience}

The Mechanics, Dynamics and Aesthetics (MDA) framework \cite{hunicke2004mda} has been immensely influential when trying to define processes and structures in the game design process. Although there have been several attempts to detail the MDA framework, including more dimensions  \cite{walk2017design}, what makes MDA so influential is its simplicity, in particular how it defines a flow where game designers establish game systems that are able to deliver experiences \cite{ham2015tabletop}. According to this perspective, designers are always uncertain about how the player will experience the game, thus requiring continuous iterations and playtesting to ensure that the game is delivering the intended experience. This effect is directly associated with the uncertainty that games provide \cite{costikyan2013uncertainty}. Zubek~\cite{zubek2020elements} highlighted the dimension of the experience when revisiting the MDA Framework \cite{hunicke2004mda,rosamonitoring}. Defining a game as an interactive system that generates experiences is not new in game studies literature. Salen and Zimmerman~\cite{salen2003rules} argue that any game fits into this definition, going beyond the anecdotal sense that games must be fun. Games can be more than this. They can be engaging activities that generate many other types of experiences since fun is subjective \cite{fullerton2014game,koster2013theory}. When addressing non-orthogames (for entertainment) \cite{elias2012characteristics}, thinking first of the experience is a way to deliver serious games, games for purposes, or applied games \cite{dorner2016serious,smith2010long}. When thinking of these types of games, do the experiences related to the serious purposes of the game fit the type of experiences that orthogames delivers to players? 
Do these serious games follow player profiles and typified experiences that proved to be engaging? Using the paths that the engagement model provides \cite{zagalo2020engagement}, are games engaging with players that like to solve problems, discover and explore the game world or allow others to be creative and relate emotionally with the characters or other players? Board games have a lower potential level of simulation than digital games. Analogue games resemble toys which automatically affect the players’ perceptions \cite{woods2012eurogames,gobet2004moves}. Can this be the reason why moving pieces that evoke child memories generates a direct play experience? Does this happen in digital games the same way, or do these tend to fall into the simulation in a way it does not feel like a playful experience? 

\subsection{Feedback}
\label{sec:multidis:feedback}

Games should be interactive \cite{fullerton2014game,ham2015tabletop,schell2008art}. Although interaction as a game concept is difficult to define, what interactivity is in practice can mean many different things. In board games, interactivity primarily pertains to the dynamic among players engaged in a multiplayer experience \cite{salen2003rules}. In video games, due to the automation of the game system, the game can react to players’ actions without player agency, providing automatic feedback. In an analogue game, the player is required to activate the game mechanisms to generate that feedback \cite{xu2011chores}. If the players do not follow the rules perfectly (or totally ignore them), the effect can be even more uncertain \cite{duarte2017distinctive}. 

Feedback, whether automated or performed by the players, is considered here as the ability of the game to provide information regarding the changes in the game state after the player performs some in-game action. Agency and its relation to feedback are thoroughly explored in game-based learning, gamification, serious games, and other applied games.

\section{The Importance of Fiction in Rehabilitation Games} 
\label{sec:fiction}

Narrative is one of the oldest ways through which humans have tried to explain the origin of the world. Some narratives, such as Hesiod’s Theogony, Homer’s Odyssey, or even the Bible, remain as long-term repositories of values and ways of interpreting the world or being in it. Through the representation mechanism, fictional narratives allow humanity to deal with possibilities, explore and test ideologies, and anticipate the effects of potential personal, social, and political choices. In this sense, fiction provides a sort of alternative, parallel life, in which the reader may safely get in contact with other ways of living \cite{saricks2001readers}.
Literary theory and criticism have long recognised the engaging power of fiction. Besides the maintenance of the debate on the cognitive value of literature \cite{huemer2022fictional,jaen2013cognitive}, the impact and the usefulness of its persuasive effect in private and public communication also challenges other fields of knowledge and operations, such as psychology, sociology, marketing and tourism. That is why storytelling appears to be, nowadays, a required competence for those who seek a job (the word ``story'' comprises a commitment to the concept of fiction that distinguishes it from factual ``history'' and evaluates the channel and the code in the communication process \cite{jakobson1960linguistics}).
Another evidence of this narrative pervasiveness is the growth of narrative games. Player communities are increasing, as well as studies on the attractive effects of narrative games that give players the power of transforming a representation of historical events, although sacrificing historical accuracy on behalf of attractiveness \cite{telles2015narrative,winnerling2014eternal}.

Despite this popularity, the field of serious games will need special attention as long as new literature concludes on the lack of knowledge about the connection of such games with narrative and/or fiction, as well as on the attractiveness of the proposed narratives \cite{sajjadi2022individualization,miller2023wrapped,hendrix2018implementing,lane2017diegetic}.
The necessity and opportunity of this exploration may be argued from the evidence, shown by literature, that storytelling and interactivity associated with cognitive training are meaningful to cognitive rehabilitation patients, and are a way to increase their interest in exercises \cite{gabele2019effects}. Besides, the impact of narratives on the experience of a learner in a serious game was revealed to rely on features such as distributed narrative, intrinsic integration (that releases pedagogical content alongside the game process), empathetic characters, and personalised narratives \cite{naul2020story}. 

\section{Conclusion: The ``Abandoned'' Artefact}
\label{sec:conclusion}

Serious games have gained significant traction in the healthcare domain. The development of a successful serious game is a multifaceted process. By providing fundamental components for the utilization of serious games in rehabilitation, this study contributes to reshaping the existing landscape in this field.
Despite the extensive body of existing research, there remains a persistent challenge in transitioning from laboratory testing to the successful implementation of a fully functional system in real-world settings. 

One of the largest roadblocks to the widespread adoption of games as an effective solution for rehabilitation is that there is no long-term proof of their effectiveness. For example, analysing the most recent top-cited surveys that consider games for stroke rehabilitation---Doumas et al.~\cite{doumas2021serious} and Mubin et al.~\cite{mubin2022exploring}---none of them consider long-term rehabilitation effectiveness in their analysis. The only consideration is the overall effectiveness, with no distinction between short- and long-term analysis. This can be problematic, considering that rehabilitation is an arduous, repetitive, and often long process, which can be mentally exhausting for patients. So it is not surprising that a game can generally be motivating when first introduced in the rehabilitation process, providing a differentiating factor. However, over time, the game may start to lose its appeal and, thus, its motivating factor.

Therefore, this study argues that one of the core requirements for such systems to be effectively adopted is to prove that motivation can be maintained throughout the rehabilitation process. Furthermore, this can vary significantly amongst clients, where recovery times may vary considerably. 
The one-size fits all approach can also impact participant motivation longevity, as tastes and preferences can diverge considerably within a population, even during rehabilitation. Different types of games (analogue and digital) could deliver an experience that would engage different users and fit a specific therapeutic approach. Game design matters to deliver the balance between playability, engagement, and serious goals. 
Effectively addressing these challenges necessitates allocating resources, including maintaining game accessibility and regular updates. A crucial consideration is whether such endeavors should be integrated into a business model associated with the entertainment industry or delivered through specialized publishers dedicated to achieving economic sustainability. This study underscores the significance of the present juncture, wherein all possibilities should be thoroughly examined, considering ethical issues and the protection of users given the critical nature of therapeutic practices.

Implementing co-creation and participatory action research at the start of serious game development is essential for enhancing research impact and facilitating multi-stakeholder engagement. This approach leverages a collaborative environment, fostering a shared commitment to ongoing and sustained stakeholder involvement. However, there is a significant paucity of completed longitudinal studies examining stakeholder engagement, underscoring the need for further research in this area. Moreover, recent studies revealed a significant gap in stakeholder involvement, especially in early design stages, despite the emphasis on customization. A detailed reporting of participatory design processes should be integrated to enhance stakeholder engagement and research reproducibility in rehabilitation \cite{rodrigues2024participation}.

Finally, our study highlights the need for a multidisciplinary group of experts to collaboratively construct shared knowledge across professional boundaries. This collaborative effort is crucial to achieve coherence and foster the integration of new perspectives in the adoption of serious games.

\section*{Acknowledgment}

This work is supported by Fundação para a Ciência e Tecnologia (FCT), HEI-Lab R\&D Unit (\url{https://doi.org/10.54499/UIDB/05380/2020}), LIBPhys R\&D Unit (\url{https://doi.org/10.54499/UIDB/04559/2020}), and UNINOVA-CTS R\&D Unit (\url{https://doi.org/10.54499/UIDB/00066/2020}). 

\bibliographystyle{ieeetr}

\end{document}